\documentclass[12pt,prl,aps]{revtex4}

\usepackage{graphicx}
\usepackage[dvips]{color}

\begin{document}

\title{Enrichment of nuclear spin isomers by\\ molecular coherent control}

\author{Pavel Chapovsky}

\affiliation{  Institute of Automation and Electrometry,
        Russian Academy of Sciences, 630090 Novosibirsk, Russia (chapovsky@iae.nsk.su)}
         

\begin{abstract}
Enrichment of nuclear spin isomers of molecules by infrared
radiation resonant to molecular rovibrational transition is considered.
Special attention is given to the enrichment by light-induced 
crossing of far separated ortho and para states.
\end{abstract}


\maketitle

\section{Introduction}

Nuclear spin isomers were discovered in the late 1920s
when ortho and para H$_2$ were separated 
by deep cooling of hydrogen gas \cite{Farkas35}. This technique was successful
because of extremely large gap between the ortho and para states
in H$_2$. Investigations of heavier isomers were not possible to perform
until recently because of the lack of appropriate enrichment methods. 
First separation of heavy molecules (CH$_3$F) was achieved
using the Light-Induce Drift effect \cite{Krasnoperov84JETPL}. Later,
the isomers of H$_2$O \cite{Konyukhov86JETPL}, H$_2$CO \cite{Kern89CPL}, 
Li$_2$ \cite{Bernheim90JCP}, and C$_2$H$_4$ \cite{Chap00CPL} were enriched.
A few practical applications of spin isomer are already known. 
For example, isomers can affect chemical 
reactions \cite{Quack77MP,Uy97PRL}, can significantly enhance NMR signals 
\cite{Bowers86PRL,Bargon93ZPC}, or can be used as spin labels in NMR based techniques.

Further investigations of nuclear spin isomers and 
progress in their applications need new 
enrichment methods. This is rather difficult and challenging task because
physical and chemical properties of spin isomers are almost identical.
Recently new approach to the isomer enrichment based on molecular excitation
by strong electromagnetic field  was proposed 
\cite{Ilichov98CPL,Shalagin99JETPL,Chap01PRA2,Chap01JPB}.  
In this paper we discuss the isomer enrichment based on the    
coherent control by strong infrared radiation. We make emphasis 
on the conditions necessary for the experimental observation of the phenomenon.  

\section{Level scheme}

\begin{figure}[thb]
\includegraphics[height=12cm]{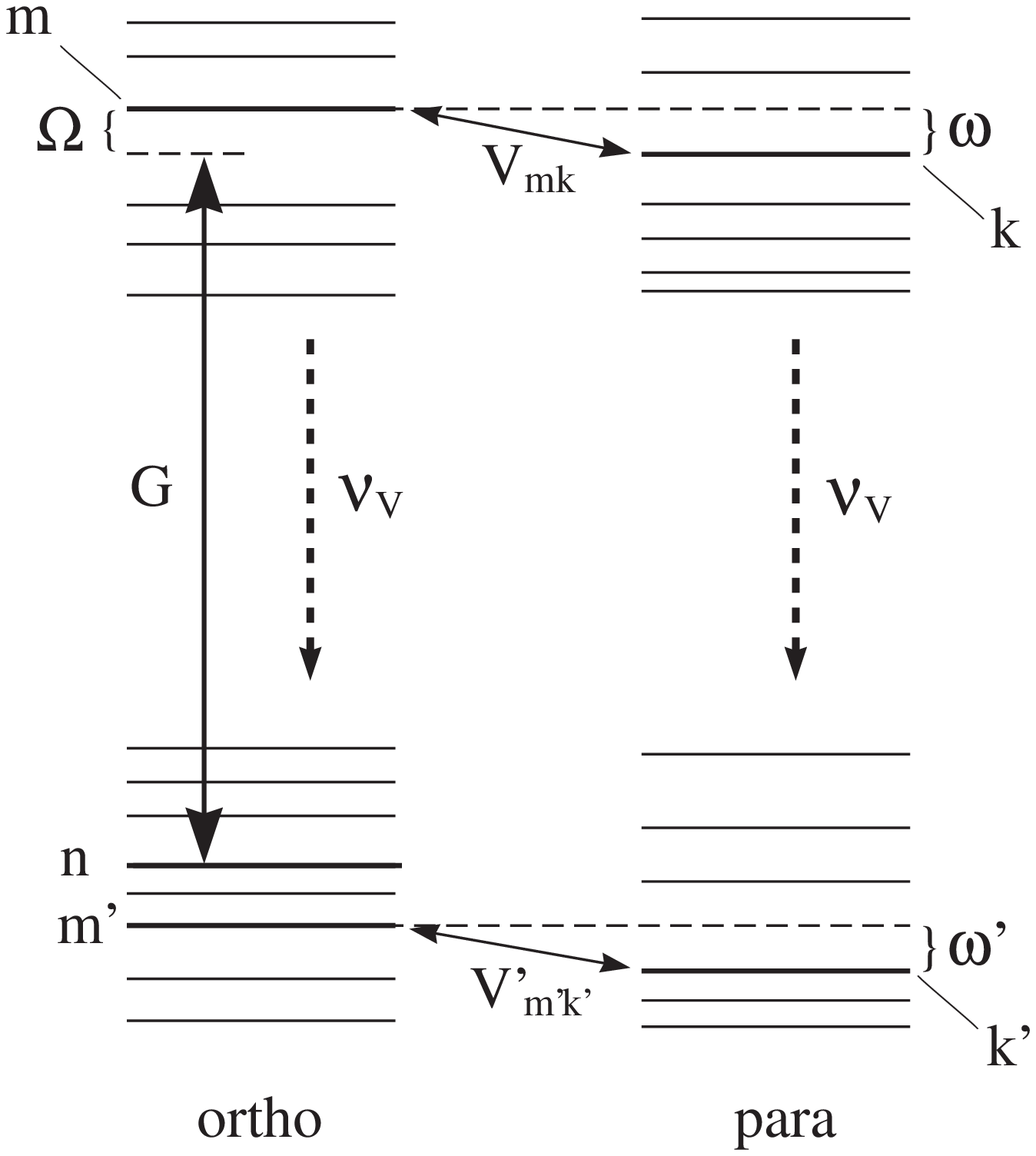}
\vspace{2cm}
\caption{Level scheme. Vertical solid line indicates infrared excitation.
Vibrational relaxation is shown by dashed vertical lines. $V$ and $V'$ 
indicate the intramolecular ortho-para state mixing.}
\label{fig1}
\end{figure}

The level scheme is shown in Fig.~\ref{fig1}.
We assume that test molecules have only two nuclear 
spin states, ortho and para. This is very often the case, 
e.g., in CH$_3$F  \cite{Landau81,Chap99ARPC}. Each 
of the two vibrational states (ground and excited) has one ortho-para 
level pair mixed by intramolecular perturbation.
Molecules are embedded into a ``nonmagnetic'' gaseous environment 
characterized by zero cross-section for collisional ortho-para transition.
The infrared radiation is resonant to the rovibrational transition 
$m-n$ of the ortho molecules.
 
The molecular Hamiltonian in our problem consists of four terms,
\begin{equation}
     \hat H = \hat H_0 + \hbar\hat H_{rad} + \hbar\hat V + \hbar\hat V'.
\label{H}
\end{equation} 
The main part, $\hat H_0$, has the eigen ortho and para states 
shown in Fig.~\ref{fig1}. $\hbar\hat H_{rad}$ describes 
the molecular interactions with the external radiation that will be
taken in the form of monochromatic travelling wave,
\begin{equation}
\hat H_{rad} = - 2\hat G\cos(\omega_Lt-{\bf kr});\ \ \ 
\hat G = {\bf E}_0\hat{\bf d}/2\hbar,   
\label{G}
\end{equation}
where ${\bf E}_0$; $\omega_L$ and ${\bf k}$ are the amplitude, frequency 
and wave vector of the electromagnetic wave, respectively; $\hat {\bf d}$ 
is the molecular electric dipole moment operator. 
$\hat V$ and $\hat V'$ are the intramolecular perturbations
that mix the ortho and para states in excited and ground vibrational
states, respectively. These perturbations originate from the nuclear
spin-spin and spin-rotation interactions 
\cite{Curl67JCP,Chap91PRA,Guskov95JETP,Ilisca98PRA}.

Suppose that a test molecule is placed 
initially in the ortho subspace of the ground vibrational state.
Due to collisions the molecule will undergo fast rotational
and velocity relaxation {\it inside} the ortho subspace. 
This will proceed until the molecule 
jumps to the state $m'$, which is mixed with the para state $k'$
by the intramolecular perturbation $\hat V'$ (see Fig.~\ref{fig1}), 
or to the state $n$ which is mixed with the para state $k$ by the 
combined action of the external field and intramolecular perturbation
$\hat V$. Admixture of a para state 
implies that the next collision can move the molecule to another para 
states and thus localize it inside the para subspace. The spin conversion
occurs. 

The molecular center-of-mass motion affects the process because of the
Doppler effect. Vibrational degree of freedom plays an important role too.
After ortho-para transition in upper vibrational state the molecule relaxes
to the ground vibrational state. Consequently, back conversion can only  
occur through the unperturbed $m'-k'$ channel. It is essential that conversion
through the $m-k$ channel can be substantially affected by electromagnetic
radiation.

Quantitative description of the isomer conversion 
can be done using the perturbation theory \cite{Chap01MP}. The zeroth-order
theory neglects the intramolecular ortho-para state mixing by 
$\hat V$ and $\hat V'$ and thus contains no ortho-para conversion. 
In this approximation an external field does not affect 
the population of para states. The  
populations of ortho states in the model of strong collisions read,
\begin{eqnarray}
 \rho_o(e,j,{\bf v})& = & \rho_o\{p(\Omega)\tau_{ev}w_jf({\bf v})+
      p(\Omega,{\bf v})[\tau_{et}w_j+\tau_e\delta_{jm}]\}, \nonumber \\
 \rho_o(g,j,{\bf v})& = & \rho_ow_jf({\bf v})-\rho_o\{
 p(\Omega)\tau_{gv}w_jf({\bf v})+p(\Omega,{\bf v})[\tau_{gt}w_j+\tau_g\delta_{jn}]\}.      
\label{sol0}
\end{eqnarray}   
Here $\rho_o$ is the total concentration of ortho molecules; 
$e,g$ designate the excited and ground vibrational states, respectively;
$j$ is the rotational quantum number; $w_j$ is the Boltzmann distribution;
$f({\bf v})$ is the Maxwell distribution and ${\bf v}$ is the molecular
center-of-mass velocity. The level populations in Eq.~(\ref{sol0}) 
are split into the terms nonequilibrium in $j$ and {\bf v}, 
nonequilibrium only in {\bf v}, and equilibrium in $j$ and {\bf v}. The
magnitudes of these terms are proportional to specific relaxation parameters 
$\tau$'s (see Ref.~\cite{Chap01MP} for more details).
The function $p(\Omega,{\bf v})$ is the excitation probability
of molecules with particular velocity {\bf v} and
$p(\Omega)=\int p(\Omega,{\bf v})d{\bf v}$, 
\begin{eqnarray}   
        Y(\Omega) & = & \int\frac{\Gamma^2_Bf({\bf v})d{\bf v}}
            {\Gamma^2_B+(\Omega-{\bf kv})^2}, \nonumber \\
        p(\Omega) & = & \frac{w_n}{\tau_1}\frac{Y(\Omega)}
            {1+\kappa^{-1}+Y(\Omega)\tau_2\tau^{-1}_1},\ \ \ 
        p(\Omega,{\bf v}) = \frac{p(\Omega)}{Y(\Omega)}
        \frac{\Gamma^2_Bf({\bf v})}{\Gamma^2_B+(\Omega-{\bf kv})^2}.
\label{ppy}
\end{eqnarray}
Here $\Omega=\omega_L-\omega$ and $\omega\equiv\omega_{mk}$ 
is the gap between the states $m$ and $k$.
The function $p(\Omega)$ is the Foigt profile of the absorption line 
having the homogeneous linewidth 
$\Gamma_B=\Gamma\sqrt{1+\kappa}$, where $\kappa$ is the saturation
parameter, $\kappa=2|G|^2\tau_1\Gamma^{-1}$; $|G|=E_0|d_{mn}|/2\hbar$ 
is the Rabi frequency;
$\Gamma$ is the collisional linewidth of the absorbing transition; 
$\tau_1=\tau_{et}w_m+\tau_{gt}w_n+\tau_e+\tau_g$ and 
$\tau_2=\tau_{ev}w_m+\tau_{gv}w_n$.

In addition to the level population change, an external radiation
creates a coherence (off-diagonal density matrix element) on the $m-n$ transition,
\begin{equation}
     \rho_o(n|m;{\bf v}) = -i\rho_o\frac{p(\Omega,{\bf v})}
     {2G\Gamma}[\Gamma-i(\Omega-{\bf kv})]e^{i(\Omega t - {\bf kr})}.
\label{ro_off}
\end{equation}

\section{Enrichment}

The first-order theory takes the intramolecular perturbations
$\hat V$ and $\hat V'$ into account. In this approximation the spin conversion
appears and it can be affected by the radiation. 
An external radiation changes   
the concentration of ortho molecules  
from the equilibrium value, $\rho_o(0)$, to the steady-state value, 
$\overline\rho_o$. Because the total concentration of the test
molecules is conserved, one has steady-state enrichment of para molecules
\cite{Chap01MP},
\begin{equation}
     \beta \equiv \frac{\overline\rho_p}{\rho_p(0)}-1 = 
     1-\frac{\gamma_{free}}{\gamma},\ \ \ 
     \gamma  =  \gamma_{free} - \gamma'_n  + \gamma_n + \gamma_{coh}.
\label{bp}
\end{equation}
Here an equal concentrations of ortho and para molecules in equilibrium
gas has been assumed that, implies $w_{m'}=w_{k'}$. The conversion 
rate $\gamma$ has four terms 
of different origin. The first one,
\begin{equation}
     \gamma_{free}=\frac{2\Gamma|V'|^2}{\Gamma^2+\omega'^2}[w_{k'}+w_{m'}], 
\label{free}
\end{equation}
is the conversion rate without an external electromagnetic field;
$\omega'\equiv\omega_{m'k'}$ is the gap between the states $m'$ and $k'$.
One can see from Eq.~(\ref{bp}) that the light-induced 
enrichment is enhanced by small $\gamma_{free}$.
The second term is due to the light-induced level depopulation in the 
ground vibrational state,
\begin{equation}
     \gamma'_n = \frac{2\Gamma|V'|^2}{\Gamma^2+\omega'^2}
     p(\Omega)w_{m'}(\tau_{gv}+\tau_{gt}) = 
     \frac{1}{2}\gamma_{free}p(\Omega)(\tau_{gv}+\tau_{gt}).
\label{g'n}
\end{equation}
The last two terms in $\gamma$ is convenient to present in a form that can be easily 
obtained from \cite{Chap01MP}. The ``noncoherent'' contribution to the rate,
\begin{equation}
\gamma_n  = \frac{2\Gamma|V|^2}{\Gamma^2+\omega^2}p(\Omega)\tau + 
            2|V|^2Re\left(1-\frac{\Gamma_1+i\omega_1}
            {\Gamma+i\omega}\right)
            \int\frac{\rho^{-1}_o\rho_o(e,m,{\bf v})d{\bf v}}
            {\Gamma_1+i(\Omega+\omega_1-{\bf kv})}.
\label{gn}
\end{equation}
Here $\tau=\tau_{ev}w_m+\tau_{et}w_m+\tau_e$. The ``coherent'' part,
$\gamma_{coh}$, originates from the light-induced coherence (\ref{ro_off}),
\begin{equation}
\gamma_{coh} = -\frac{|V|^2p(\Omega)}{\Gamma^2+\omega^2}+ 
                \frac{|V|^2}{\Gamma}Re\frac{\Gamma+\Gamma_1+i\omega_1}
                {\Gamma+i\omega}\int\frac{p(\Omega,{\bf v})d{\bf v}}
                {\Gamma_1+i(\Omega+\omega_1-{\bf kv})}.
\label{gcoh}
\end{equation}
The new homogeneous width, $\Gamma_1$, and the ortho-para level gap, $\omega_1$, are 
\begin{equation}
     \Gamma_1 = \Gamma\left(1+\frac{|G|^2}{\Gamma^2+\omega^2}\right),\ \ \
     \omega_1 = \omega\left(1-\frac{|G|^2}{\Gamma^2+\omega^2}\right). 
\label{G1}
\end{equation}

One can see from Eqs.~(\ref{g'n}), (\ref{gn}), and
(\ref{gcoh}) that the conversion rate, $\gamma$, has two peaks. The peak  
at the absorption line center ($\Omega=0$) appears because
the excitation probabilities, $p(\Omega)$ and $p(\Omega,{\bf v})$, 
have maximum at $\Omega=0$. The homogeneous width
of this peak is equal to $\Gamma_B$ that can be very large in molecular systems
due to the slow rotational, velocity, and vibrational relaxations. The peak
at $\Omega=-\omega_1$ is due to the light-induced crossing of $m$ and $k$ 
states (see Ref.~\cite{Chap01PRA2,Chap01JPB,Chap01MP}) 
and appears formally because of the resonant denominators in the integrands in
Eq.~(\ref{gn}) and (\ref{gcoh}). The homogeneous width of the peak at 
$\Omega=-\omega_1$ is equal to $\Gamma_1$.

Coherent control allows to achieve substantial
enrichment if the mixing in ground and vibrationally excited states
has the same order of magnitude \cite{Chap01MP}. On the other hand, 
close coincidences between ortho and para states in molecules are rather rare.
Consequently, such close level pairs in excited vibrational state
may be absent, or they cannot be accessible by the available radiation. 
Below we estimate the enrichment in the case of large ortho-para
gap in excited state, $\omega\gg kv_0\gg\Gamma$, where $kv_0$ is 
the Doppler parameter and $v_0$ is the molecular thermal velocity, 
$v_0=\sqrt{2k_BT/M}$. For moderate radiation intensities, $|G|\ll\omega$, 
the field-dependent terms at the absorption line center, $\Omega=0$, are
\begin{equation}
     \gamma'_n(0) \simeq 0.5\gamma_{free}p(0)(\tau_{gv}+\tau_{gt});\ \ \ 
     \gamma_n(0)  \simeq \frac{2|V|^2}{\omega^2}p(0)\tau\Gamma;\ \ \ 
     \gamma_{coh}(0) \simeq -\frac{|V|^2}{\omega^2}p(0). 
\label{gat0}
\end{equation}
The terms $\gamma_n(0)$ and $\gamma_{coh}(0)$ have the ``off-resonant'' nature
which manifests itself by the large ortho-para gap, $\omega$, in 
the denominators. At large $\omega$, the 
conversion rate, $\gamma(0)$, is determined mainly by $\gamma'_n(0)$
and the radiation depletes the para molecules,
\begin{equation} 
\beta(0)\simeq-0.5p(0)(\tau_{gv}+\tau_{gt}).
\label{b0}
\end{equation}
This is analogous to the enrichment by light-induced population change 
\cite{Ilichov98CPL,Shalagin99JETPL}.
Maximum of $\beta(0)$ is achieved at sufficiently strong field, $\Gamma_B\gg kv_0$. 
In this case, $Y(0)\simeq1$, $p(0)\simeq w_n/(\tau_1+\tau_2)$ and enrichment 
reaches the value,
\begin{equation}
     \beta_{max}(0) \simeq -w_n\frac{\tau_{gv}+\tau_{gt}}{\tau_1+\tau_2}.
\label{bmax}
\end{equation} 
Depending on the relaxation scheme and the origin of the relaxation
``bottle neck'', one can achieve different level of enrichment. In
the most favorable situation of fast velocity and rotational relaxations 
the enrichment reaches -1/4.
 
Larger enrichment can be achieved at the level-crossing resonance.
The field-dependent rates at the $\Omega=-\omega_1$ can be readily estimated
if $\Gamma_B\gg kv_0$. In this case
$p(-\omega_1,{\bf v})\simeq p(-\omega_1)f({\bf v})$ and one has,
\begin{eqnarray}
     \gamma'_n(-\omega_1) &\simeq& 0.5\gamma_{free}p(-\omega_1)
             (\tau_{gv}+\tau_{gt}), \nonumber \\
     \gamma_n(-\omega_1) &\simeq& \frac{2\sqrt{\pi}|V|^2}{kv_0}\frac{|G|^2}{\omega^2}
              p(-\omega_1)\tau ,\ \ \ 
     \gamma_{coh}(-\omega_1) \simeq \frac{\sqrt{\pi}|V|^2}{kv_0}p(-\omega_1)\Gamma^{-1}.
\label{gat1}
\end{eqnarray}
The terms, $\gamma_n(-\omega_1)$ and $\gamma_{coh}(-\omega_1)$ are due to the
light-induced crossing of the $m$ and $k$ states that has an efficient width equal 
$kv_0/\sqrt{\pi}$ (compare
with the level crossing in molecules being at rest \cite{Chap01JPB}). 
The main contribution to $\gamma(-\omega_1)$ comes from $\gamma_{coh}(-\omega_1)$
if $|G|\ll\omega$.

The results (\ref{gat0}) and (\ref{gat1}) allows to determine which peak,
$\Omega=0$, or $\Omega=-\omega$, should be used for enrichment. At low
radiation intensity larger enrichment is achieved at $\Omega=0$ because
$p(0)\gg p(-\omega_1)$ if $\Gamma_B\ll\omega$.
If radiation intensity is high, $\Gamma_B\sim\omega$, larger enrichment is 
obtained at $\Omega=-\omega_1$. The ratio of the two peaks reads,
\begin{equation}
     \frac{\gamma(-\omega_1)}{\gamma(0)} \simeq \frac{\gamma_{coh}(-\omega_1)}{\gamma'_n(0)} 
     \sim \frac{2\sqrt{\pi}|V|^2}{\gamma_{free}kv_0}\Gamma^{-1}(\tau_{gv}+\tau_{gt})^{-1}. 
\label{ratio}
\end{equation}

\begin{figure}[thb]
\includegraphics[height=12cm]{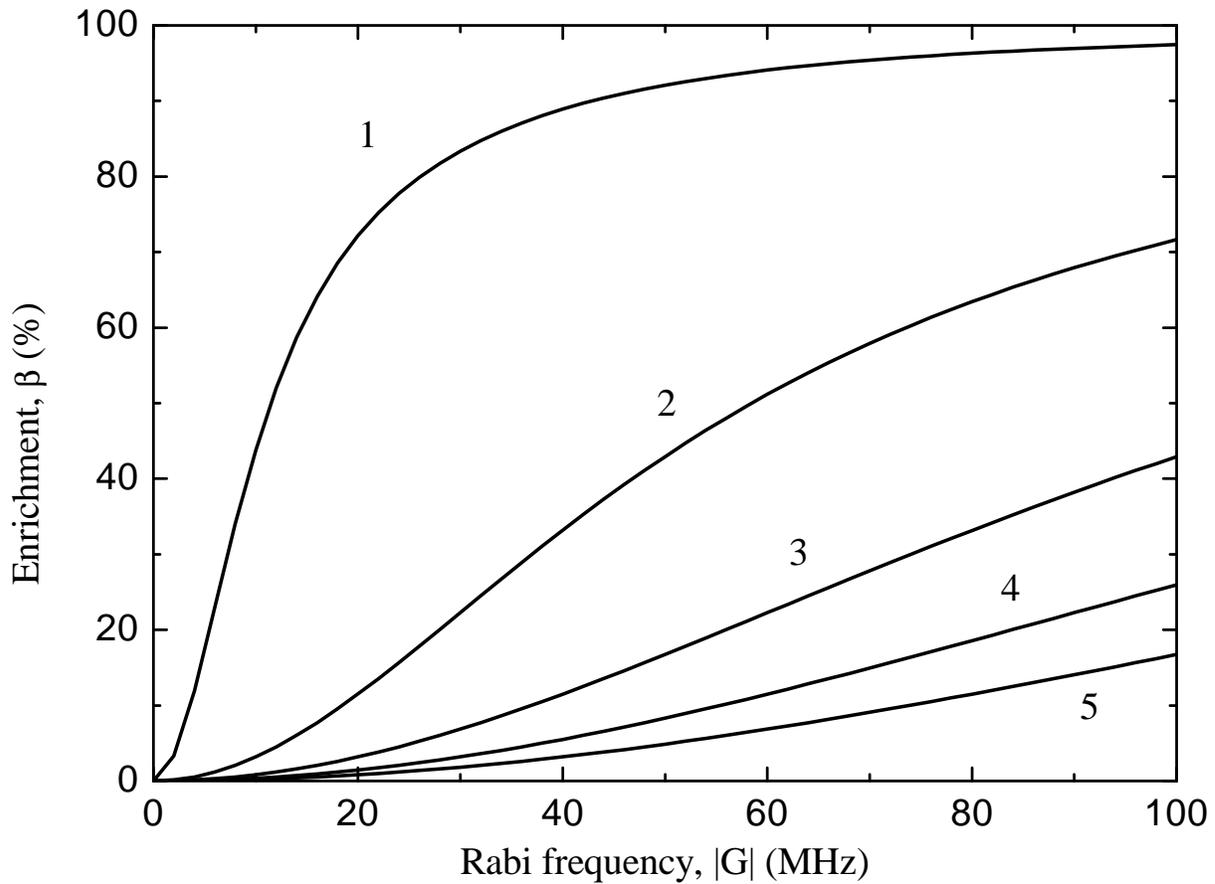}
\vspace{2cm}
\caption{Enrichment of para molecules as a function of Rabi frequency,
$|G|$, and various ortho-para level gaps in excited vibrational state, 
$\omega$. 1) $\omega=100$~MHz, 2) 500~MHz, 3) 1000~MHz, 4) 1500~MHz, 5) 2000~MHz.}
\label{fig2}
\end{figure}

The amplitude of the enrichment peak at $\Omega=-\omega_1$ as a function 
of $|G|$ and various values of $\omega$ are shown in Fig.~\ref{fig2}. The 
calculations were done using the exact expressions for the conversion rates,
(\ref{free})--(\ref{gcoh}). In this
numerical example the following parameters were used \cite{Chap01MP}, 
$V=3$~kHz, $V'=5$~kHz, $\Gamma=2$~MHz, $\omega'=130$~MHz, $kv_0=30$~MHz;
$\tau_1=5$~MHz$^{-1}$; $\tau_2=1$~MHz$^{-1}$ and all Boltzmann factors
equal, $w_m=w_n=w_k=w_{m'}=w_{k'}=10^{-2}$. 

\section{Discussion}

Conditions necessary for the experimental observation of the light-induced enrichment
are not easy to fulfil. First of all,  the field free conversion rate,
$\gamma_{free}$, of the molecules under consideration should be low in order to
enhance the role of light-induced changes in the total rate, $\gamma$. 
Note that the rate $\gamma_{free}$ should combine all significant contributions, e.g.,
caused by the molecular collisions with the walls of the gas container. At present,
only in a few rare cases (see the Introduction) the rates $\gamma_{free}$ are
known.

Next task is to determine the target ortho-para level pair in the excited vibrational
state. For this, one has to know rather accurately the relative positions of ortho and para states
and the intramolecular ortho-para mixing efficiency as well.

Perhaps, the most difficult problem is to find powerful radiation
having right frequency. The most promising sources of such radiation
are the powerful Free Electron Lasers. 

\section*{Acknowledgments}

This work was made possible by financial support from the  
Russian Foundation for Basic Research (RFBR), grant No.~01-03-32905
and from The Institute of Physical and Chemical Research (RIKEN), 
Japan.

\newpage

\newpage

\end{document}